\documentclass[prb,superscriptaddress,amsfonts,amssymb,amsmath,floats,twocolumn,aps,footinbib,shownopacs]{revtex4}

\usepackage[pdftex]{graphicx}
\usepackage{subfigure}
\usepackage{dsfont}
\usepackage{lipsum}
\usepackage{dcolumn}
\usepackage{bm}
\usepackage{color}
\usepackage{physics}
\usepackage{multirow}
\usepackage[pdftex,colorlinks=true, linkcolor=blue,citecolor=blue,filecolor=blue]{hyperref}
\newcommand{\bi}{\begin{itemize}}
\newcommand{\ei}{\end{itemize}}
\newcommand{\be}{\begin{equation}}
\newcommand{\ee}{\end{equation}}
\newcommand{\ba}{\begin{eqnarray}}
\newcommand{\ea}{\end{eqnarray}}

\newcommand{\imag}{\mathrm{i}}

\newcommand{\dt}{\delta_t}

\begin{document}

\title{Memory truncated Kadanoff-Baym equations}

\author{Christopher Stahl}
\affiliation{Department of Physics, University of Erlangen-Nuremberg, 91058 Erlangen, Germany}

\author{Nagamalleswararao Dasari} 
\affiliation{Department of Physics, University of Erlangen-Nuremberg, 91058 Erlangen, Germany}
\affiliation{I. Institute of Theoretical Physics, Department of Physics, University of Hamburg, 22607 Hamburg, Germany}
\author{Jiajun Li} 
\affiliation{Paul Scherrer Institute, 5232 Villigen, Switzerland}

\author{Antonio Picano} 
\affiliation{Department of Physics, University of Erlangen-Nuremberg, 91058 Erlangen, Germany}
\author{Philipp Werner}
\affiliation{Department of Physics, University of Fribourg, 1700 Fribourg, Switzerland}
\author{Martin Eckstein}
\affiliation{Department of Physics, University of Erlangen-Nuremberg, 91058 Erlangen, Germany}

\begin{abstract}
The Keldysh formalism for nonequilibrium Green's functions is a powerful theoretical framework for the description of the electronic structure, spectroscopy, and dynamics of strongly correlated systems. However, the underlying Kadanoff-Baym equations (KBE) for the two-time Keldysh Green's functions involve a memory kernel which results in a high computational cost for long simulation times $t_\text{max}$, with a cubic scaling of the computation time with $t_\text{max}$.  Truncation of the memory kernel can reduce the computational cost to linear scaling with $t_\text{max}$, but the required memory times will depend on the model and the diagrammatic approximation to the self-energy. 
We explain how a  truncation of the memory kernel can be incorporated into the time-propagation algorithm to solve the KBE, and investigate the systematic truncation of the memory kernel for the Hubbard model in different parameter regimes, and for different diagrammatic approximations. The truncation is easier to control within dynamical mean-field solutions, where it is applied to a momentum-independent self-energy. Here, simulation times up to two orders of magnitude longer are accessible both in the weak and strong coupling regime, allowing for a study of  long-time phenomena such as the crossover between pre-thermalization and thermalization dynamics. 
\end{abstract} 
\maketitle 

\section{\label{sec:Introduction}Introduction}

Femtosecond pump-probe experiments on complex solids \cite{Giannetti2016} as well as quantum simulations with cold gases \cite{Senaratne2018,Chiu2018,GuardadoSanchez2018} allow to investigate collective quantum behavior on microscopic timescales. The dynamics of such many-particle systems often involves processes on vastly different timescales, which is a particular challenge for theoretical simulations. For example, both the femtosecond electron dynamics and the evolution of the  order parameter on the picosecond scale are relevant for the non-thermal evolution of symmetry broken phases in correlated electron systems. The phenomenological  time-dependent Ginzburg-Landau theory can describe the dynamics of the order parameter on coarse-grained time-scales  
\cite{Lemonik2017, Lemonik2018, Lemonik2018a, Dolgirev2020, maklar2020nonequilibrium,Grandi2021}, 
but the link to a microscopic theory which can treat non-thermal electrons and collective degrees of freedom on equal footing is not straightforward \cite{Stahl2021}. 

Field theoretical techniques based on non-equilibrium Green's functions~(NEGF) and the Keldysh-formalism \cite{kamenev2011field, stefanucci2013nonequilibrium, aoki2014} 
can provide a versatile realistic description of non-equilibrium experiments involving transport~\cite{schlunzen2016},  dynamics  in quantum gases \cite{Schluenzen2017,sandholzer2019}, or pump-probe experiments in correlated solids~\cite{ligges2018,gillmeister2020,li2018theory, golez2017,kumar2019higgs,sentef2013}. Within this formalism, a major task is to obtain the Green's functions $G(t,t')$ with their dependence on two time arguments from the many-body self-energy  $\Sigma(t,t')$ via the Kadanoff-Baym equations~(KBE). The latter are in essence equations of motion for $G$, in which $\Sigma$ acts as a memory kernel. The main numerical cost is given by the evaluation of memory integrals, i.e., the convolution of $G$ and $\Sigma$ over earlier times. For an equidistant time discretization with $N_t$ steps, the computational effort and the required computer memory shows a cubic scaling $\mathcal{O}(N_t^3)$ and a quadratic scaling $\mathcal{O}(N_t^2)$ with $N_t$, respectively, which is the bottleneck  for many applications.  

There are various paths to overcome this bottleneck. One possibility is to approximate the full propagation scheme using the generalized Kadanoff Baym ansatz~\cite{lipavsky1986,kalvova2019}. This can eventually give a linear scaling  $\mathcal{O}(N_t)$ of the computational effort \cite{schlunzen2020} and works well in many situations~\cite{schuler2020,murakami2020ultrafast,schuler2019, karlsson2020, tuovinen2020comparing}. However, the accuracy of the underlying approximations, in particular the use of Hartree Fock propagators for the spectral function and the ansatz itself, are rigorously justified at weak coupling and have to be evaluated case by case. An alternative approximation strategy are quantum kinetic equations 
\cite{kadanoff1962quantum,Stark2013, Bonitz2016,Kremo1997,Wais2018,Schutzhold2019,haug2008quantum,picano2021},
which rely on the gradient expansion and are justified in particular when there are well separated timescales. In addition to these approximate strategies, one can look for an efficient numerical solution of the full KBE. Larger times can be accessed using parallel implementations~\cite{balzer2012nonequilibrium,talarico19}, and high-order time stepping and quadrature rules~\cite{schuler2020} which reduce the number of time-steps. Beyond this brute force approach, a promising direction are compressed storage representations of the two-time Green's functions which are compact in memory but can nevertheless be incorporated into a time-stepping procedure with little computational overhead \cite{Kaye2021}. 

A conceptually simpler approach is to truncate the memory integrals in time. In many physical situations, the self-energy $\Sigma(t,t')$  decays to zero at large time differences, so that a truncation is possible with a controlled error. In Ref.~\onlinecite{schuler2018}, this truncation has been explored for various situations, in particular related to simulations based on non-equilibrium dynamical mean-field theory (DMFT), by simply setting the kernel to zero at large time difference  in an existing implementation of the full KBE. This provided a proof-of-principles for this approach, but the simulations were limited to times which are also accessible within the solution of the full KBE. Here we describe how the memory truncation can be incorporated into a time-stepping approach to yield a linear scaling $\mathcal{O}(N_t)$ of the computational effort. An implementation of this approach has already been used in combination with non-equilibrium DMFT simulations both at weak \cite{Picano2020} and strong \cite{Dasari2021} coupling, where two orders of magnitude longer times could be accessed, compared with the non-truncated implementation. The purpose of the present paper is to describe the technical aspects of the formalism, and present additional test cases which analyze paradigmatic problems with a large separation of timescales: Prethermalization and thermalization after quenches in the Hubbard model, the thermalization of a large gap Mott Insulator, and the dynamics of symmetry broken states. The first two examples are based on non-equilibrium DMFT \cite{aoki2014}, where only local (momentum-averaged) self-energies and Green's functions are truncated. The last example uses the perturbative fluctuation exchange formalism to construct the self-energy, so that  the truncation is applied to momentum-dependent self-energies and susceptibilities, making the choice of the memory cutoff more subtle.

 The paper is organized as follows: Sec.~\ref{sec:TruncKBE} presents a discussion of the Kadanoff-Baym equation and the truncation scheme in the weak and strong coupling limit. In Sec.~\ref{sec:TruncLocal} we demonstrate the implementation of the truncated KBE  for local self-energies in the Hubbard model.  The following Sec.~\ref{sec:FLEX} is dedicated to the discussion of the truncation scheme for the momentum-dependent self-energy and RPA equations within the fluctuation-exchange approximation for the dynamics of superconducting fluctuations in the attractive Hubbard model. Sec.~\ref{sec:Conclusion} contains a discussion and outlook.

\section{\label{sec:TruncKBE}Formalism}

\subsection{Kadanoff-Baym equations}

We consider a system described by the general Hamiltonian
\begin{align}
\label{Hgeneral}
H = \sum_{\alpha,\alpha'}\epsilon_{\alpha,\alpha'}(t) \, c_{\alpha}^\dagger c^{\mathstrut}_{\alpha'} + H_{\rm int}.
\end{align}
Here $c_{\alpha}^\dagger$ and $c^{\mathstrut}_{\alpha}$ are fermionic creation and annihilation operators for single-particle states $\alpha$, which can label spin, momentum, and orbital. The first term denotes a general time-dependent non-interacting Hamiltonian in terms of the matrix $\epsilon$, the second term is the interaction, which will be specified for the various applications below. In the Keldysh formalism, one solves the nonequilibrium dynamics in terms of the contour-ordered single-particle Green's functions $G_{\alpha,\alpha'}(t,t')=-\imag\langle T_\mathcal{C}c_{\alpha}(t)c_{\alpha'}^\dagger(t')\rangle$. 
The subscript $\alpha$ will be omitted in the following, and the Green's function is understood to represent a matrix in orbital space.
Here the time-arguments are located on the $L$-shaped Keldysh contour, with a forward and backward branch for times $t>0$ on the real axis, and the imaginary branch from $0$ to $-\imag\beta$  where $\beta$ is the inverse temperature of the initial equilibrium state. $T_{\mathcal{C}}$ denotes the corresponding time-ordering operator (see, e.g.,  Ref.~\onlinecite{stefanucci2013nonequilibrium} or \onlinecite{aoki2014} for a review of the formalism; the notation mainly follows Ref.~\onlinecite{aoki2014}.) 

The interacting Green's function $G$ is expressed in terms of the many-body self-energy $\Sigma(t,t')$ through the Dyson equation
\begin{align}
\label{inetraga}
G=&\,G_0+G_0\ast\Sigma\ast G,
\end{align}
where $\ast$ denotes a convolution in time on the Keldysh contour and the matrix multiplication in the orbital space, and $G_0$ is the non-interacting Green's function. With the inverse of $G_0$ on the contour $\mathcal{C}$,
\begin{align}
G_0^{-1}(t,t')= \delta_{\mathcal{C}}(t,t')[\imag\partial_t-\epsilon(t)],
\end{align}
Eq.~\eqref{inetraga} becomes an integro-differential equation 
\begin{equation}
[\imag\partial_t-\epsilon(t)]G(t,t')-\int_{\mathcal{C}}\!\dd \bar{t}\,\,\Sigma(t,\bar{t})G(\bar{t},t')=\delta_{\mathcal{C}}(t,t').
\label{eqn:KeldyshA}
\end{equation}
Here $\delta_{\mathcal{C}}(t,t')$ represents the Dirac delta function on the Keldysh contour, and $\epsilon(t)$ the single-particle Hamiltonian of Eq.~\eqref{Hgeneral}. Given an approximation to the self-energy $\Sigma$, the main task for the solution of the non-equilibrium problem is the solution of Eq.~\eqref{eqn:KeldyshA}. For this purpose, $G$ and $\Sigma$ are parametrized in terms of independent functions with real and imaginary time arguments. For a general two-time function $X$ on $\mathcal{C}$, we will use the notation
\begin{align}
X^<(t,t') &= X(t_{+},t_{-}),
\\
X^>(t,t') &= X(t_{-},t_{+}),
\\
X^R(t,t') &= \theta(t-t') [X^>(t,t')-X^<(t,t')],
\\
X^A(t,t') &= \theta(t'-t) [X^<(t,t')-X^>(t,t')],
\\
X^\rceil(t,\tau) &= X(t_{\pm},-\imag\tau),
\\
X^\lceil(t,\tau) &= X(-\imag\tau,t_{\pm}),
\\
X^M(\tau-\tau') &= -\imag X(-\imag\tau,-\imag\tau'),
\end{align}
where $t_{\pm}$ denotes time $t$ on the upper/lower Keldysh contour, and $-\imag\tau$ is an argument on the imaginary time branch. These functions are partially redundant; following the numerical scheme described in Ref.~\onlinecite{schuler2020}, we will use the retarded $(R)$, lesser $(<)$, left-mixing $(\rceil)$ and Matsubara ($M$) component to parametrize integral equations on $\mathcal{C}$. In addition, unless indicated otherwise there is a hermitian symmetry,
\begin{align}
X^>(t,t') &= -X^>(t',t)^\dagger,
\\
X^<(t,t') &= -X^<(t',t)^\dagger,
\label{symles}
\\
X^A(t,t') &= X^R(t',t)^\dagger,
\\
X^\rceil(t,\tau) &= \mp X^\lceil(\beta-\tau,t)^\dagger.
\end{align}
The upper/lower sign in the last equation corresponds to bosonic and fermionic Green's functions, respectively.
When written in terms of the retarded, lesser, mixing, and Matsubara components, Eq.~\eqref{eqn:KeldyshA} reads
\begin{equation}
[\imag\partial_{t}-\epsilon(t)]G^R(t,t')-\int_{t'}^t\!\!\! \dd \bar{t}\,\,\Sigma^R(t,\bar{t})G^R(\bar{t},t')=0,
\label{eqn:KBEretarded}
\end{equation}
\begin{align}
[\imag\partial_t -\epsilon(t)]G^\rceil(t,\tau)
&-\int_0^t\!\!\!\dd \bar{t}\,\,\Sigma^R(t,\bar{t})G^\rceil(\bar{t},\tau)
\label{eqn:KBEleft-mixing}\\
&=\int_0^\beta\!\!\! \dd \tau'\,\, \Sigma^\rceil(t,\tau')G^M(\tau'-\tau),\notag
\end{align}
\begin{align}
[\imag\partial_t-\epsilon(t)]G^<(t,t')
-\int_0^t\!\!\! \dd \bar{t}\,\,\Sigma^R(t,\bar{t})G^<(\bar{t},t')\label{eqn:KBElesser}\\
=\int_0^{t'}\!\!\! \dd \bar{t}\,\,\Sigma^<(t,\bar{t})G^R(t',\bar{t})^\dagger\notag
\pm\imag\int_0^\beta\!\!\! \dd \tau\,\, \Sigma^\rceil(t,\tau')&
G^\rceil(t',\beta-\tau')^\dagger.\notag
\end{align}
The first equation is subject to the initial condition $G^R(t,t)=-\imag$. In addition, there is a separate equation for the Matsubara component $G^M$ in terms of $\Sigma^M$, which is not relevant for the discussion of the real time evaluation below. The derivation of Eqs.~\eqref{eqn:KBEretarded}-\eqref{eqn:KBElesser}, using Langreth rules to rewrite the convolution, is given in the literature \cite{aoki2014, stefanucci2013nonequilibrium} and will not be reproduced here. In the following, we will discuss how these equations are solved with a memory-truncated kernel. 

For the most straightforward solution, the time axis is discretized with a constant time step $\dt$, so that Eqs.~\eqref{eqn:KBEretarded}-\eqref{eqn:KBElesser} can be solved in a time-stepping procedure: 
Assuming that all functions have been determined at real time arguments $m\dt$ for $m\le n-1$, one can use Eq.~\eqref{eqn:KBEretarded} to extend the solution of $G^R$ to time step $n$, i.e., determine $G^R(n\dt,m\dt)$, for all $m\le n$, because the integral in Eq.~\eqref{eqn:KBEretarded} depends only on $G^R(t,t')$ at $t,t'\le n\dt$. Next, because the integrals in Eq.~\eqref{eqn:KBEleft-mixing} depend only on the previously computed $G^R$ as well as on $G^\rceil(m\dt,\tau)$ for $m\le n$, one can use Eq.~\eqref{eqn:KBEleft-mixing} to determine $G^\rceil(nh,\tau)$ at time step $n$ (for all $\tau$). Finally, Eq.~\eqref{eqn:KBElesser} is used to determine $G^<(n\dt,m\dt)$ at time step $n$, for all $m\le n$. 

\begin{figure}
\centerline{\includegraphics[width=0.8\columnwidth]{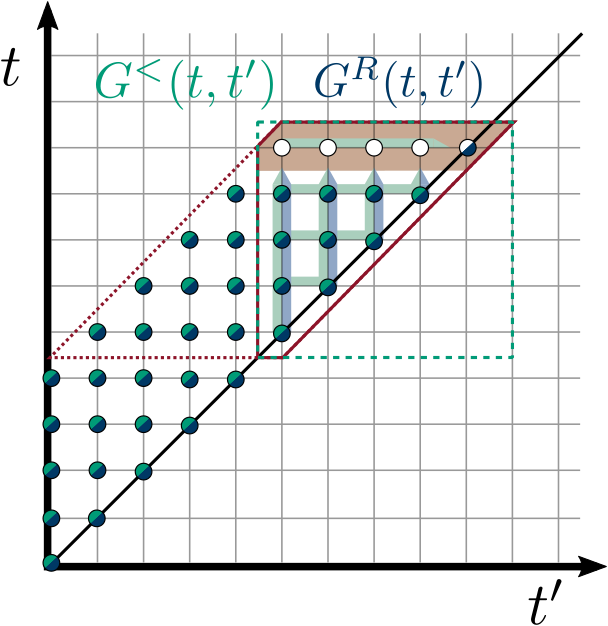}}
\caption{Truncated evaluation scheme for the 
KBE, Eqs.~\eqref{eqn:KBE_retarded_trunc} and \eqref{eqn:KBE_lesser_trunc}. The shaded region indicates the time slice \eqref{eqn:Time_Slice}. The solid red contour indicates the moving triangular window~\eqref{eqn:Time_tri_window}, which contains all data points of the  retarded~(blue) and lesser~(green) Green's function at previous time steps that are needed to solve the KBE at the shaded time step, as indicated by the arrows. Note that the required values of the lesser functions enclosed by the dashed green contour can be constructed from \eqref{symles}. The red dashed contour indicates the extended moving window (see text).}
\label{fig:KBEscheme}
\end{figure}

\subsection{Truncation scheme}

The computational effort for the solution of Eqs.~\eqref{eqn:KBEretarded} to \eqref{eqn:KBElesser}, as described above, scales like $\mathcal{O}(n^2)$ at time step $n$, and thus like $\mathcal{O}(n^3)$ to propagate to time $n\dt$. However, because temporal correlations tend to decay, the memory integrals in the KBE can be truncated. For example, for a system coupled to a reservoir one  would expect that the Green's function itself shows an exponential long time-decay $G^{R,<}(t,t')\sim e^{-\eta|t-t'|}$ and $G^\rceil(t,\tau)\sim e^{-\eta t}$, where $\eta$ is determined by the bandwidth of the bath and the temperature. In other situations, the Green's function may not rapidly decay, but the self-energy does. For example, the decay of momentum resolved Green's functions $G_{\bm k}$ for $\bm k$ in a Fermi liquid is set by the  quasiparticle lifetime, which is peaked at the Fermi surface, while the self-energy has a much weaker dependence on $\bm k$. It is therefore a natural starting point to consider $\Sigma(t,t')=0$ for time differences beyond a cutoff, i.e.,
\begin{align}
\begin{array}{l}
\Sigma^R(t,t')=\Sigma^<(t,t')=0
\\
\Sigma^\rceil(t,\tau)=0
\end{array}
\text{~~for~~}
\begin{array}{l}
\abs{t-t'}>t_c,
\\
t>t_c.
\end{array}
\label{eqn:TruncConstraint}
\end{align}
The memory cutoff $t_c$ will depend on the physical problem, and can later be treated as a numerical control parameter. 
When Eq.~\eqref{eqn:TruncConstraint}  is used within Eqs.~\eqref{eqn:KBEretarded} to \eqref{eqn:KBElesser}, one finds that
the left-mixing component is now decoupled from the lesser and retarded component for $t>t_c$ in Eq.~\eqref{eqn:KBElesser}, and the equations reduce to
\begin{equation}
[\imag\partial_t-\epsilon(t)]G^R(t,t-s)-\int_{t-s}^t\!\!\!\dd \bar{t}\,\,\Sigma^R(t,\bar{t})G^R(\bar{t},t-s)=0
\label{eqn:KBE_retarded_trunc}
\end{equation}
for the retarded component and
\begin{gather}
[\imag\partial_t-\epsilon(t)]G^<(t,t-s)
-\int_{t-t_c}^t\!\!\!\dd \bar{t}\,\,\Sigma^R(t,\bar{t})G^<(\bar{t},t-s)\notag\\
=\int_{t-t_c}^{t'}\!\!\!\dd{\bar{t}}\Sigma^<(t,\bar{t})G^R(t-s,\bar{t})^\dagger
\label{eqn:KBE_lesser_trunc}
\end{gather}
for the lesser component. These equations now allow to determine the Green's function $G(t,t')$ on the domain $|t-t'|<t_c$ from the self-energy on the same domain. To make this explicit we define the partial time-slice  
\begin{align}
\mathcal{T}[X]_n^{n_c}\equiv\{&X^R(n\dt,n\dt-m\dt),\label{eqn:Time_Slice}
\\&X^<(n\dt,n\dt-m\dt),0\leq m\leq n_c\},\notag
\end{align}
for any contour function $X$ on the equidistant mesh ($t_c=n_c\dt$ ), as shown in Fig.~\ref{fig:KBEscheme} by the red shaded region.  From Eq.~\eqref{eqn:KBE_retarded_trunc}, we can confirm that the retarded component of $\mathcal{T}[G]^{n_c}_n$ can be  calculated from $G^R(t_1,t_2)$ and $\Sigma^R(t_1,t_2)$ in the domain $t-t_c\leq t_1 \leq t$ and $t-t_c\leq t_2\leq t_1$, which is the triangular domain shown by the solid red line in Fig.~\ref{fig:KBEscheme}.  In the following, we denote this triangular domain by
\begin{equation}
\mathcal{\bar{M}}[X]_n^{n_c}\equiv\bigcup_{m=0}^{n_c}\mathcal{T}[X]_{n-m}^{n_c-m}.
\label{eqn:Time_tri_window}
\end{equation}
The equivalent analysis of Eq.~\eqref{eqn:KBE_lesser_trunc} yields that the memory integrals can be solved if both retarded and lesser components of $G$ and $\Sigma$ are given on this triangular window. Note that $G^<$ is in principle needed on the green dashed rectangular domain, but the lower triangular of the latter can be reconstructed from the upper triangular using the hermitian symmetry \eqref{symles}. We can thus set up an implicit time-stepping function which calculates $\mathcal{T}[G]_n^{n_c}$ based on $\bar{\mathcal{M}}[\Sigma]_n^{n_c}$ and $\bar{\mathcal{M}}[G]_n^{n_c}$. 
The numerical implementation uses the same quadrature and differentiation as described for the full KBE in Ref.~\onlinecite{schuler2020}. 
After solving for $\mathcal{T}[G]_n^{n_c}$, one can then shift the moving window to $\mathcal{M}[G]_{n+1}^{n_c}$, where now only $\mathcal{T}[G]_{n+1}^{n_c}$ is unknown, and proceed to solve for time step $n+1$. The solution on the first $n_c$ time steps is obtained by solving the full KBE equation, including the imaginary branch of $\mathcal{C}$. The computational complexity of the truncated KBE evaluation scheme is therefore given by $\mathcal{O}(nn_c^2)$, while the memory complexity is $\mathcal{O}(n_c^2)$, effectively bypassing the limiting memory bottleneck of the full evaluation scheme. For a more efficient memory allocation it is convenient to keep the redundant parallelogrammatic moving window in memory, as shown by the red dotted contour in Fig.~\ref{fig:KBEscheme}.

Before testing the truncated KBE in various contexts, let us make some comments:
(i)  An analogous time propagation based on a moving two-time window can be set up for other integral equations on $\mathcal{C}$, most notably the equation
\begin{equation}
G + K\ast G = Q,
\label{vie2}
\end{equation}
with known inhomogeneity $Q$ and kernel $K$,  which has to be solved in the context of an RPA formalism (see Sec.~\ref{sec:FLEX}).
(ii) Often, the self-energy $\Sigma$, or the kernel $K$ in Eq.~\eqref{vie2} is itself a functional of $G$, and the solution of the Dyson equation  has to be supplemented by an iterative calculation of $\Sigma$ or $K$. A self-consistent propagation scheme can be set up if  the self energy on a given time step $n$,  $\mathcal{T}[\Sigma]_n^{n_c}$, requires only knowledge of $\mathcal{T}[G]_n^{n_c}$, or can be approximated in that way. Whether this is the case or not depends on the system of interest, but it will hold for all applications studied below.
(iii) As already mentioned, if the self-energy satisfies Eq.~\eqref{eqn:TruncConstraint}, the result for $G^R(t,t')$ and $G^<(t,t')$ at $|t-t'|<t_c$ is exact, even if the Green's function does not vanish for $|t-t'|>t_c$. Knowledge of $G^R(t,t')$ and $G^<(t,t')$ on the truncated window $|t-t'|<t_c$ allows to construct all equal-time observables, such as the orbital occupation $n_\alpha(t) = -\imag G^<_{\alpha\alpha}(t,t)$. Also relevant two-particle quantities like the interaction energy can be obtained from the convolution $\Sigma\ast G$. Spectral information, which is related to the Fourier transform of $G^R(t,t')$ and $G^<(t,t')$ as a function of $t-t'$, can therefore only be obtained  with a limited frequency resolution $\sim 1/t_c$ (see also discussion of the specific examples below). To compute $G^R(t,t')$ at $|t-t'|>t_c$, one could use the  equation conjugate to Eq.~\eqref{eqn:KBEretarded},
\begin{equation}
-\imag\partial_{t'}G^R(t,t')=G^R(t,t')\epsilon(t') +\int_{t'}^t\!\!\! \dd \bar{t}\,\,G^R(t,\bar{t})\Sigma^R(\bar{t},t'),
\label{eqn:KBEretardedconj}
\end{equation}
and propagate $t'$ backward at fixed $t$. This would require to keep $\Sigma$ in memory in the whole domain $|t-t'|<t_c$, and therefore increase the memory  requirement  at time $t=n\dt$ from  $\mathcal{O}(n_c^2)$ to $\mathcal{O}(nn_c)$, still smaller than $\mathcal{O}(n^2)$ as for the full KBE.

\subsection{Non-equilibrium DMFT}

Nonequilibrium DMFT is an extension of DMFT \cite{georges1996} to the Keldysh framework, which has been applied in various ways to study the dynamics of strongly correlated electrons \cite{aoki2014}.  The solution of nonequilibrium DMFT requires the solution of multiple integral equations of the type \eqref{eqn:KeldyshA} or  \eqref{vie2}. In the following section, we briefly review the formalism in order to explain how the memory truncation can be applied in this context. Two paradigmatic applications at weak and strong coupling will be given in Sec.~\ref{sec:TruncLocal}.

Within DMFT, a lattice problem with local interaction is mapped to an Anderson impurity model, defined by the action
\begin{align}
S=&\int_{\mathcal{C}} dt H_{\rm imp}(t) +\int_{\mathcal{C}} dtdt' \sum_{\alpha\alpha'}c_{\alpha}^\dag(t)\Delta_{\alpha\alpha'}(t,t')c_{\alpha'}(t').
\label{eqn:Anderson_model}
\end{align}
This action describes a single site with local Hamiltonian $H_{\rm imp}$, coupled to a reservoir of free electrons. The local Hamiltonian $H_{\rm imp}$ contains the electron interaction on the impurity, while the hybridization function $\Delta_{\alpha\alpha'}(t,t')$ is obtained by integrating out the reservoir degrees of freedom \cite{aoki2014}. The impurity model is used to obtain the local Green's function $G_{\rm imp}$ and the impurity self-energy $\Sigma_{\rm imp}$, which are related by the Dyson equation
\begin{align}
\label{dysonimp}
[ \imag\partial_t -\epsilon_{\rm imp} - \Sigma_{\rm imp} - \Delta] \ast G_{\rm imp}= \delta_{\mathcal{C}},
\end{align}
where $\epsilon_{\rm imp}$ is the single particle contribution in $H_{\rm imp}$. The impurity self-energy is used as an approximation for the lattice self-energy, so that the lattice Green's functions at momentum $\bm k$ are obtained as
\begin{align}
\label{eqdyson}
[\imag\partial_t -\epsilon_{\bm k} - \Sigma_{\rm imp}]\ast G_{\bm k} = \delta_{\mathcal{C}},
\end{align}
and the DMFT equations are closed by requiring the self-consistency relation
\begin{align}
\label{glof}
G_{loc} \equiv \sum_{\bm k} G_{\bm k} \stackrel{!}{=}G_{\rm imp}.
\end{align}
Here the sum is assumed to be normalized, $\sum_{\bm k}=1$. 
A useful way to close the self-consistent equations explicitly is to recast Eqs.~\eqref{dysonimp} to \eqref{glof} into an integral equation for  $\Delta$  in terms of $G_{\bm k}$,
\begin{align}
\label{EqforDelta}
\Delta + G_1 \ast \Delta = G_2,
\end{align}
where $G_1=\sum_{\bm k} \epsilon_{\bm k} G_{\bm k}$, $G_2=\sum_{\bm k}\epsilon_{\bm k} + \sum_{\bm k} \epsilon_{\bm k} G_{\bm k} \epsilon_{\bm k} $. 

The explicit implementation of the DMFT equations depends on how the impurity model is solved. For the case of the strong coupling expansion this will be described separately in Sec.~\ref{sec:StrongScheme}. In a weak-coupling expansion, $\Sigma_{\rm imp}$ is obtained as a series either in $G_{\rm imp}$, or in the non-interacting impurity Green's function $\mathcal{G}_{\rm imp}$, 
e.g., using the iterated perturbation theory (IPT)
\begin{align}
\label{sigma2}
\Sigma_{\rm imp}(t,t') = -U^2 \mathcal{G}_{\rm imp}(t,t')\mathcal{G}_{\rm imp}(t,t')\mathcal{G}_{\rm imp}(t',t).
\end{align}
The  non-interacting impurity Green's function in turn is given by the solution of
\begin{align}
\label{weiss}
[\imag\partial_t -\epsilon_{\rm imp}- \Delta]\ast \mathcal{G}_{\rm imp} = \delta_{\mathcal{C}}.
\end{align}
The resulting self-consistent equations \eqref{weiss}, \eqref{sigma2}, \eqref{eqdyson},  and \eqref{EqforDelta} can be solved in a moving time window $|t-t'|<t_c$, provided that the kernels $\Sigma_{\rm imp}$, $\Delta$, and $G_1$ can be truncated accordingly. In practice, one would solve all equations with a  given  $t_c$, and increase $t_c$ until convergence. An example will be presented in Sec.~\ref{sec:TruncLocal}, where we will also discuss the relative data quality for the truncation of the various kernels. 

\subsection{\label{sec:StrongScheme}Strong-coupling Approximation}
\newcommand\ct{{%
  \ooalign{$\circlearrowright$\cr
    \hidewidth$*$\hidewidth}}}

The strong-coupling expansion of the DMFT impurity problem \eqref{eqn:Anderson_model} is suitable when the interaction term dominates  over the hybridization, such as in Mott insulators.\cite{Keitler,coleman1984}  The impurity Green's function is then obtained by an expansion in the hybridization function $\Delta$. The memory truncation can be used naturally in Eq.~\eqref{EqforDelta} for the determination of $\Delta$, but also within the hybridization expansion itself. The purpose of the following section is to explain how the memory truncation scheme is used in the perturbative strong-coupling expansion for the Anderson impurity model. For a more detailed  derivation of the real-time hybridization expansion, see Refs.~\onlinecite{eckstein2010} and~\onlinecite{aoki2014}.

\begin{figure}
\includegraphics[width=\columnwidth]{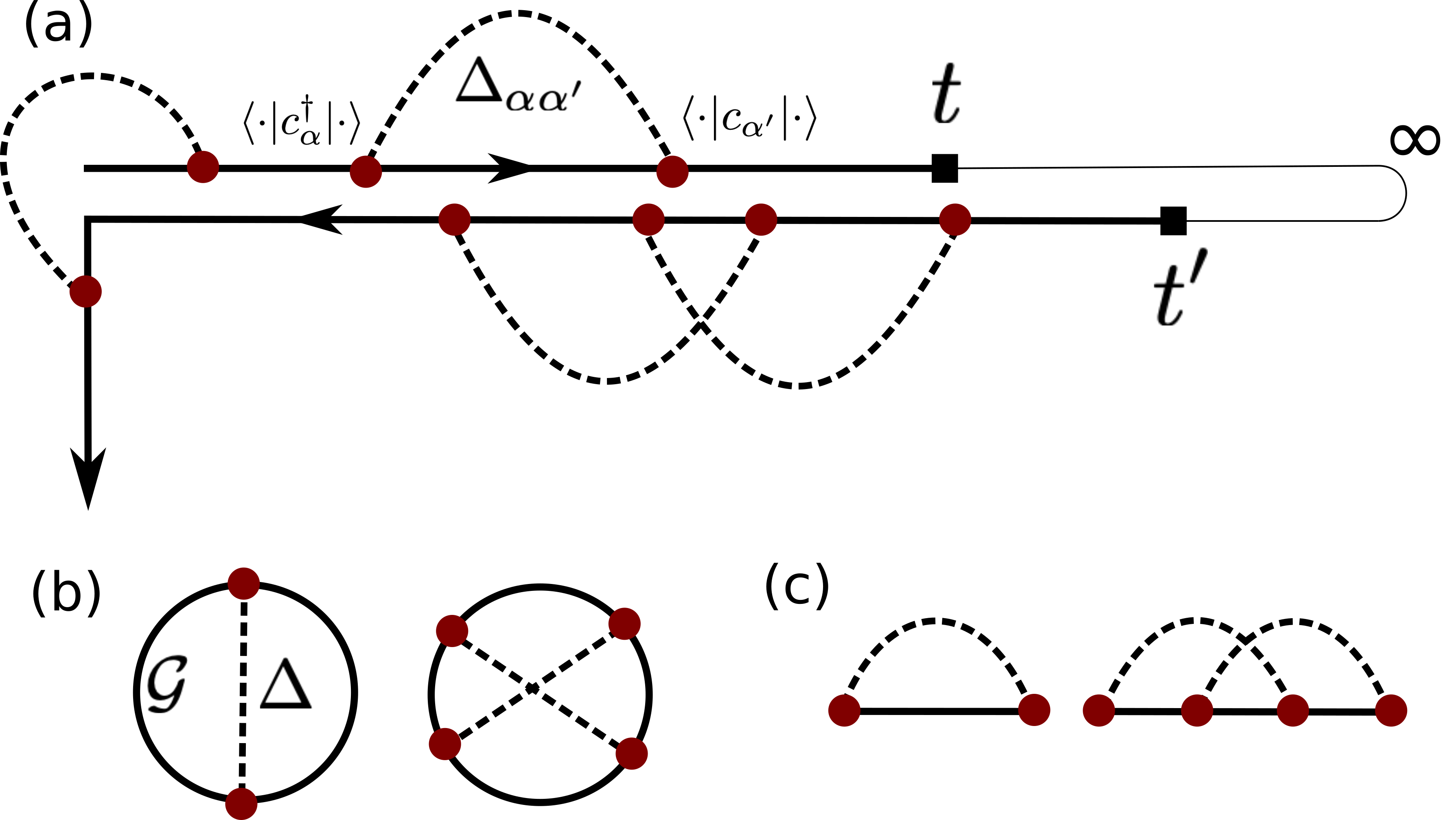}
\caption{The strong-coupling expansion on the Kadanoff-Baym contour. The solid line represents a propagator, while the dashed lines are hybridization functions. The red dot indicates an interaction vertex, i.e., the matrix element of $c_\alpha,c^\dag_{\alpha}$ in the eigenbasis of $H_{\rm imp}$. 
(a) A contribution to the dressed propagator $\mathcal{G}(t,t')$ 
is obtained by inserting hybridization lines $\Delta_{\alpha\alpha'}$ on the time contour. 
The retarded (lesser) Green's function corresponds to $t<t'$ on the lower (upper) contour. 
(b) The Luttinger-Ward functional for the pseudoparticle Green's functions, to second order in $\Delta$. 
(c) NCA and OCA self-energies, obtained from the diagrams in (b). 
}
\label{strong_coupl}
\end{figure}

The building blocks of the strong-coupling perturbation series are the pseudoparticle retarded and lesser Green's functions, $\mathcal{G}^{R}(t,t')$ and $\mathcal{G}^{<}(t,t')$. They are defined as matrix elements of the time-evolution operator $\bra{n}e^{-iH_{\rm imp}(t-t')}\ket{m}$ on eigenstates $\ket{n},\ket{m}$ of the local impurity problem, dressed by an arbitrary number of bath hybridization lines $\Delta(t,t')$, see Fig.~\ref{strong_coupl}. 
The propagators can be used to construct the partition function of the Anderson impurity model and to evaluate physical observables, in particular the physical impurity Green's function. The propagators $\mathcal{G}$ satisfy similar equations of motion as the physical Green's functions \cite{eckstein2010}, with a pseudoparticle self-energy $\Sigma^{R,<}$ that can be systematically generated from  a Luttinger-Ward functional. The truncated equations of motion read
\begin{align}
\label{eqn:pp_eom_ret}
(\imag\partial_t-H_{\rm imp} ) &\mathcal{G}^R(t,t')-\int^t_{t-t_c}\!\!\! \dd\bar{t}\,\,\Sigma^R(t,\bar{t}) \mathcal{G}^R(\bar{t},t')=0, \\
\label{eqn:pp_eom_les}
(\imag\partial_t-H_{\rm imp} ) &\mathcal{G}^<(t,t')-\int^t_{t-t_c}\!\!\! \dd\bar{t}\,\,\Sigma^R(t,\bar{t}) \mathcal{G}^<(\bar{t},t')\nonumber\\
&=\quad\quad\int^{t'}_{t'-t_c}\!\!\! \dd\bar{t}\,\,\Sigma^<(t,\bar{t}) \mathcal{G}^A(\bar{t},t'), 
\end{align}
where the advanced propagator $\mathcal{G}^A(t,t')=[\mathcal{G}^R(t',t)]^\dag$. These equations have the same causal structure as those in the weak-coupling theory~(\ref{eqn:KBE_retarded_trunc}, \ref{eqn:KBE_lesser_trunc}).
In this article, we will concentrate on the lowest-order diagrams for the self-energy $\Sigma$, the non-crossing approximation (NCA). For the truncated time regime, the imaginary time axis can be neglected as $t,t'\gtrsim t_c$. The self-energy is then given by 
\begin{align}
\label{eqn:nca}
\Sigma^R_{nm}(t,t')&=\imag\sum_{n'm'}\mathcal{G}^R_{n'm'}(t,t')[c_{\alpha,nn'}\Delta^>_{\alpha\alpha'}(t,t')c^\dag_{\alpha,m'm}\nonumber\\
&\quad - c^\dag_{\alpha,nn'}\Delta^>_{\alpha\alpha'}(t',t)c_{\alpha,m'm}],\nonumber\\
\Sigma^<_{nm}(t,t')&=\imag\sum_{n'm'}\mathcal{G}^<_{n'm'}(t,t')[c_{\alpha,nn'}\Delta^<_{\alpha\alpha'}(t,t')c^\dag_{\alpha,m'm}\nonumber\\
&\quad-c^\dag_{\alpha,nn'}\Delta^<_{\alpha\alpha'}(t',t)c_{\alpha,m'm}],
\end{align}
where we have defined the shorthand $c_{\alpha,nm}=\bra{n} c_{\alpha} \ket{m}$ for the electronic operator $c_\alpha$. Under the same approximation, the physical Green's function can be evaluated by inserting creation and annihilation operators of physical fermions on the Kadanoff-Baym contour \cite{eckstein2010}, 
\begin{align}
G^{\lessgtr}_{\alpha\alpha'}(t,t') &= \imag\sum_{\{m\}}(-1)^{m_1} c_{\alpha, m_2 m_1}c^\dag_{\alpha',m'_1 m'_2}\,\,\,\,\nonumber\\
&\times\quad\mathcal{G}_{m_2 m'_2}^{\lessgtr}(t,t')\mathcal{G}^{\gtrless}_{m_1'm_1}(t',t)/Z,
\end{align}
where the summation $\{m\}$ is over all $m_i$'s,
and $Z=\imag\sum_m(-1)^m\mathcal{G}^<_{mm}(t,t)$ is the normalization factor.
This is essentially the first diagram of Fig.~\ref{strong_coupl}(b) with the hybridization line removed.

By combining Eq.~\eqref{eqn:nca} with Eqs.~\eqref{eqn:pp_eom_ret} and \eqref{eqn:pp_eom_les}, we obtain a closed set of equations, which can be numerically solved from time $t_0$ given the initial values of propagators for $t\in[t_0-t_c,t_0]$. The detailed numerical implementation based on the moving windows $\mathcal{M}[\mathcal{G}]^{n_c}_n$ and $\mathcal{M}[\Sigma]^{n_c}_n$ is  analogous to the weak-coupling theory. It is worth noting that the 
decay of $\Sigma^{R,<}(t,t')$ is determined by the hybridization function $\Delta^{\lessgtr}(t,t')$ in this case.
The dressed propagators $\mathcal{G}^{R,<}(t,t')$ 
can decay slowly for large time separations $|t-t'|$, particularly at low temperatures \cite{coleman1984, menge1988}, but nevertheless,  one has $\Sigma^{R,<}(t,t')\approx 0$  for $|t-t'|\gtrsim t_c$ provided $\Delta^{\lessgtr}(t,t')\approx 0 $. Therefore, the truncation of  $\Sigma$  can still be justified  in the equation of motion Eqs.~\eqref{eqn:pp_eom_ret} and \eqref{eqn:pp_eom_les}, as long as the hybridization function decays fast enough. We will numerically confirm this scenario in Sec.~\ref{sec:NCA}. 

\section{\label{sec:TruncLocal}Truncation of local approximation schemes}
\subsection{\label{sec:QuenchHub}Interaction quench in the Hubbard model}

\begin{figure}[tbp]
\includegraphics[width=0.5\textwidth]{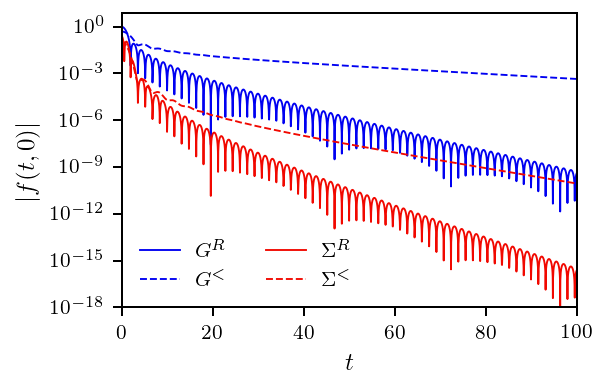}
\caption{\label{fig:PM_AFM_truncation}
Absolute value of the self-energy $\Sigma$~(red) and the local Green's function $G$~(blue), for lesser~(dashed) and retarded~(solid) components, as a function of relative time $|t-t'|$ for the quench $U:0 \to 1$.}
\end{figure}

\begin{figure}[tbp]
\includegraphics[width=0.5\textwidth]{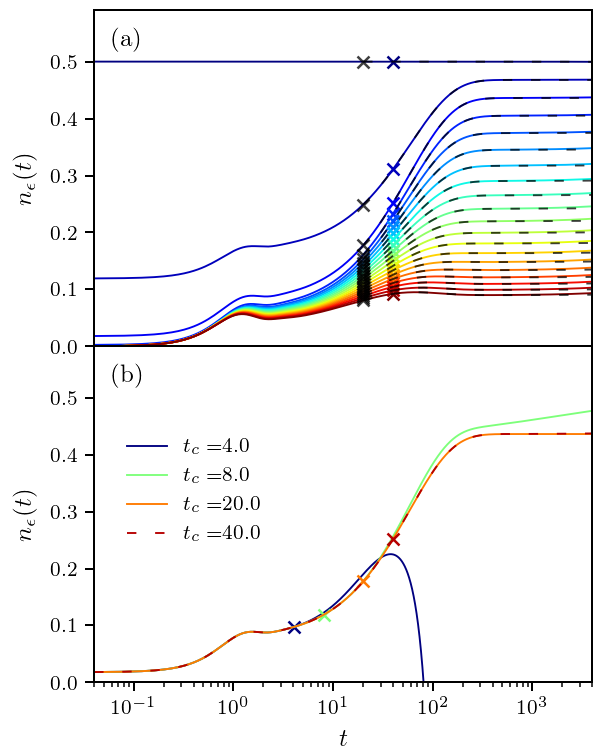}
\caption{Quench $U:0 \to 1$. (a) The curves show the momentum occupation $n_\epsilon(t)$ for single-particle energies $\epsilon=0,0.02,...,0.38$ (blue to red) above the Fermi energy and within the  half-bandwidth of $W/2$=2. Colored (black) curves represent calculations  at cutoff time  $t_c=40$ ($t_c=20$) which is indicated by the marker $X$. (b) Convergence analysis for $\epsilon=0.04$ at different cutoff times $t_c=4,8,20,40$ (blue, green, orange, red line), with marker $X$ indicating the cutoff time.
}
\label{fig:KBE_eps}
\end{figure}

As a first example for the memory truncation within the solution of the KBE, we investigate an interaction quench in the Hubbard model:
\begin{align}
	\label{eqn:Hubbard_pm}
	H = -t_h \sum_{\langle i,j \rangle,\sigma} c_{i\sigma}^\dagger c_{j\sigma} + U(t)\sum_{j} n_{j\uparrow}n_{j\downarrow}+ \mu \sum_{i,\sigma} n_{i,\sigma}.
\end{align}
Here $c_{j\sigma}$ ($c_{i\sigma}^\dagger$) denotes the annihilation (creation) operator for a fermion with spin $\sigma\in\{\uparrow,\downarrow\}$ on lattice site $j$ ($i$); $n_{j\sigma}=c_{j\sigma}^\dagger c_{j\sigma}$ is the number of particles with spin $\sigma$ on lattice site $j$. The Hubbard model~\eqref{eqn:Hubbard_pm} represents fermions which hop with amplitude $t_h$ between nearest neighbor sites $\langle i,j\rangle$ on a given lattice, and interact with a local time-dependent repulsion $U(t)$; $\mu$ is the chemical potential. The hopping amplitude is fixed to $t_h=1$ setting the energy and time scales.

We explore the dynamics following a quench from the non-interacting system $U=0$ to the interacting one at $U=1$, associated with a paramagnetic state. After the quench, the dynamics is expected to show a fast pre-thermalization of the electronic distribution, followed by a very slow thermalization~\cite{Moeckel2008,eckstein2009,Stark2013}. We solve the model~\eqref{eqn:Hubbard_pm} on a Bethe lattice, with a semi-elliptic density of states $D_0(\epsilon)=\sqrt{4 t_h^2-\epsilon^2}/(2\pi t_h^2)$, 
using DMFT with a self-consistent second-order impurity solver. 
This implies a closed form self-consistency relation $\Delta=t_h^2G$ for the hybridization function.
The equations to be solved are therefore the impurity Dyson equation \eqref{dysonimp}, with $\epsilon_{\rm imp}=0$ for the half filled case and the second-order self-energy  $\Sigma(t,t')=U(t)G(t,t')G(t,t')G(t',t)U(t')$.

The untruncated solution on the first one hundred inverse hoppings  in Fig.~\ref{fig:PM_AFM_truncation} shows a rapid decay of $\Sigma^R(t,0)$ and $G^R(t,0)$. The lesser component of the self-energy  decays on a similar time-scale, while the lesser component of the Green's function maintains values above $10^{-3}$. The slow decay of $G^<(t,0)$ is related to the sharp drop in the distribution function at low temperatures, which also characterizes the pre-thermalized state after the excitation of the system. 
After truncating the relevant integral kernels $\Delta$ and $\Sigma$, the explicit integral equations for the lesser and retarded 
components of Eq.~\eqref{dysonimp} read
\begin{align}
\label{eqn:localG_ret}
\imag\partial_t G^R(t,t')-&\int^t_{t-t_c}\!\!\! d\bar{t}\,\,[\Sigma^R+\Delta^R](t,\bar{t}) G^R(\bar{t},t')=0, \\
\label{eqn:localG_les}
\imag\partial_t G^<(t,t')-&\int^t_{t-t_c}\!\!\! d\bar{t}\,\,[\Sigma^R+\Delta^R](t,\bar{t}) G^<(\bar{t},t')=\nonumber\\
&\quad\quad\int^{t'}_{t'-t_c}\!\!\! d\bar{t}\,\,[\Sigma^<+\Delta^<](t,\bar{t}) G^A(\bar{t},t'). 
\end{align}
The system is therefore well suited for the application of the truncation scheme despite the slow decay of the lesser component of the hybridization function, because the latter enters only via a convolution with the more rapidly decaying $G^A(t,t')=G^R(t',t)^\dagger$ so that converged results can be expected for reasonable cutoff times $t_c$.

To discuss the physical behavior we calculate the momentum-dependent Green's function $G_{\bm k}$. Within DMFT, $G_{\bm k}$ depends on $\bm k$ only via $\epsilon_{\bm k}=\epsilon$. The corresponding KBE for $G_{\bm k,\sigma}\equiv G_\epsilon$ therefore reads
\begin{align}
&[\imag\partial_t -\epsilon ] G_{\epsilon}(t,t')
&-\int_{\mathcal{C}}\! d\bar t\,\, \Sigma(t,\bar t) G_{\epsilon}(\bar t, t') =\delta(t,t').
\label{eqn:KBE_eps}
\end{align}
The truncation of the integral kernel $\Sigma$ in this equation is also possible, as evident from Fig.~\ref{fig:PM_AFM_truncation}.
The solution of Eq.~\eqref{eqn:KBE_eps} yields the evolution of the momentum occupation $n_\epsilon(t)=\imag G_\epsilon^<(t,t)$, plotted in Fig.~\ref{fig:KBE_eps} for different energies
above the Fermi energy.
The occupation of the initial state 
corresponds to a step-like Fermi-distribution at $T=0.01$. 
After the interaction quench the momentum occupation for energies above the Fermi energy increases rapidly and reaches a plateau around $t=1$. 
\begin{figure}[tbp]
\includegraphics[scale=1]{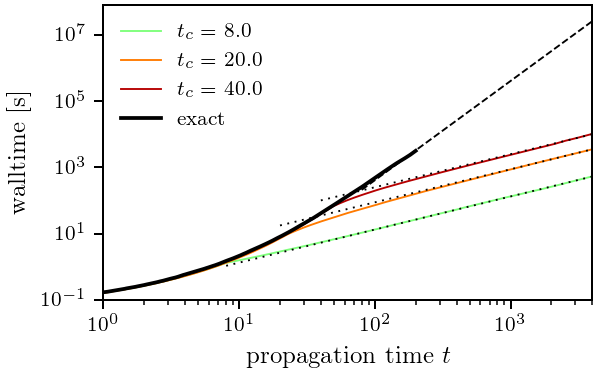}
\caption{\label{fig:Scaling} Computation time on a single processor for the DMFT simulations shown in Fig.~\ref{fig:KBE_eps} [Eq. \eqref{eqn:localG_ret}-\eqref{eqn:localG_les}], for cutoff times $t_c = 8, 20, 40$ (green, orange, red), and the non-truncated solution (black). The dashed line shows the asymptotic $\mathcal{O}(n^3)$ scaling of the non-truncated solution, while the dotted lines correspond to a linear scaling $\mathcal{O}(n)$ of the calculation time with the propagation time $t$.}
\end{figure}
The dynamics of this fast pre-thermalization is followed by
a slow thermalization within several hundred inverse hoppings. 
The dependence of the results on the cutoff time $t_c$ is analyzed in Fig.~\ref{fig:KBE_eps}(b). One finds
a convergence for $t_c>20$, and already qualitatively correct results for $t_c=8$. Only for the extreme limit $t_c\leq4$, corresponding to $n_c=100$ points on the memory grid, an instability of the equations was observed, resulting in non-physical solutions. As demonstrated in Fig.~\ref{fig:Scaling}, the computation time scales like $\mathcal{O}(n)$ after $t_c$, in contrast to the cubic scaling $\mathcal{O}(n^3)$ of the non-truncated solution. 

Because $\Sigma$ decays faster than $\Delta$, one could, instead of solving the Dyson equation for $G_{imp}$ with a large cutoff $n_c$, solve explicitly the momentum dependent Dyson equation \eqref{eqn:KBE_eps} for a suitable set of $N_k$  momenta with a shorter memory cutoff $n_{c,\Sigma}$, and construct $G_{loc}$ from Eq.~\eqref{glof}.  The total memory and computation time then scales like $\mathcal{O}(N_k n_{c,\Sigma}^2)$ and $\mathcal{O}(N_k n n_{c,\Sigma}^2)$, respectively. (Note that the problem can be parallelized over the momenta).  Which of the two approaches is optimal depends on the problem; for the DMFT solution of the Hubbard model on the Bethe lattice in the antiferromagnetic phase a particularly slow decay of $\Delta$ made using the momentum dependent Dyson equation beneficial. \cite{Picano2020}

\subsection{\label{sec:NCA}Impact ionization in the Mott-Insulator}

\begin{figure}[tbp]
\includegraphics[scale=1]{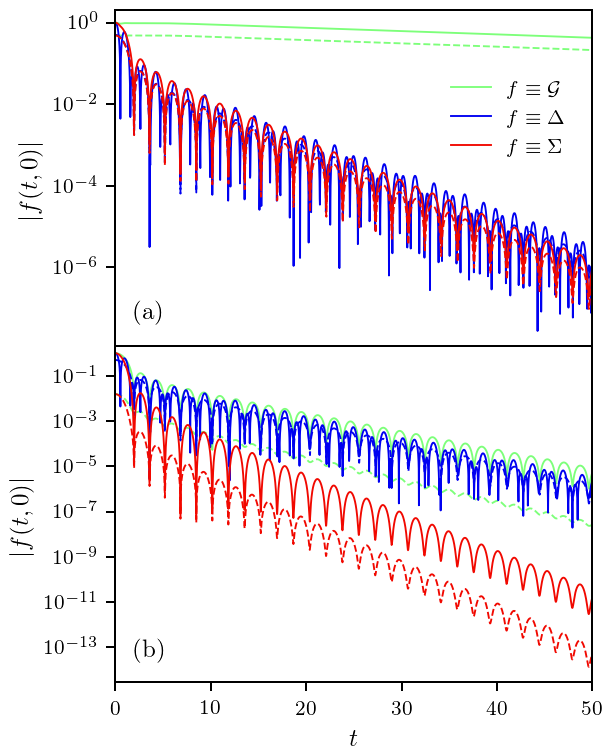}
\caption{\label{fig:NCA_FIG1}
(a) Absolute value of the retarded (solid) and lesser (dashed) component of the hybridization function $\Delta$ (blue), the pseudoparticle Green's function $\mathcal{G}_{mm}$ (green), and the self-energy $\Sigma_{mm}$ (red) for the singly occupied state ($m=\uparrow,\downarrow$). Results are shown for the Hubbard interaction $U$=5.7 and frequency of the laser pulse $\Omega=3.5\pi/2$. (b) Same quantities as in (a), but for $m=0,\uparrow\downarrow$ (which are identical because of particle-hole symmetry).
}
\end{figure}

As second example in the context of DMFT we study the thermalization of a photo-excited Mott insulator. This explores the strong coupling limit of the Hubbard model~\eqref{eqn:Hubbard_pm}, where $U$ exceeds the bandwidth $W$.  An impulsive excitation of the Mott insulator can lead to the creation of 
doublons (doubly occupied sites in the Mott insulator) and empty sites, which becomes evident through occupation in the upper Hubbard band.  If the excess energy of the excited particles is larger than the band gap, a conversion of high energy doublons to low energy doublons can occur via the excitation of additional doublon-hole pairs across the Mott gap. Such an ``impact ionization" has been previously studied in Ref.~\onlinecite{Werner2014}, where the discussion was limited to the early stages of the dynamic. The application of the truncation scheme allows the extension of the numerical simulations by two orders of magnitude, so that the full thermalization of the electronic system becomes accessible.

Starting from the equilibrium paramagnetic Mott insulating state at $\beta=10$, the photo-excitation by a laser field with vector potential $A(t)$ is incorporated by a Peierls approximation, which gives a complex hopping amplitude $t_h \exp(-\imag\phi(t))$ with $\phi(t)= e a A(t)/\hbar c$. The relation between the vector potential $A(t)$ and the laser field $E(t)$ is given by the integral $A(t)=-\int^t_0\dd sE(s)$. We use Gaussian pulses of the form 
\begin{equation}
E(t)=E_0e^{-(t-t_p)^2/{\delta}^2}\sin(\Omega(t-t_p)),
\end{equation}
with duration $\delta^2$=6, centred around time $t_p$=6 ($e$,$a$,$\hbar$, $c$ are set to unity). The frequency $\Omega$ will be varied to modify the initial energy distribution of the photo-excited doublons, while the amplitude of the pulse $E_0$ determines the excitation density. The model is solved on a Bethe lattice with bandwidth $W=4$, using DMFT and the NCA solver as explained in Sec.~\ref{sec:StrongScheme}. With the Peierls phase, the closed form self-consistency for the hybridization function is modified to (see, e.g., Ref.~\onlinecite{Dasari2021})
\begin{equation}
\Delta(t,t')=\frac{t_h^{2}}{2} 
\sum_{\eta=\pm}
e^{\eta\imag\phi(t)}G(t,t')e^{-\eta\imag\phi(t')},
\end{equation}
which is in essence an average of the hybridization with neighboring sites in the direction of the fields ($\eta=+$) and against the field ($\eta=-$). The impurity hybridization function in the paramagnetic phase decays rapidly  below a threshold of $10^{-5}$ on the first 50 inverse hopping times, as shown in Fig.~\ref{fig:NCA_FIG1}. Following Eq.~\eqref{eqn:nca}, this leads to a fast decay of the pseudo-particle self-energy $\Sigma_{mm}$ both for the 
singly occupied state ($m=\uparrow,\downarrow$) and for empty or doubly occupied sites ($m=0,\uparrow\downarrow$)~(see Fig.~\ref{fig:NCA_FIG1}), even though the pseudoparticle Green's function $\mathcal{G}_{mm}$ for the singly occupied state decays very slowly. As the integral kernel in the strong coupling approximation (Eqs.~\eqref{eqn:pp_eom_ret} and \eqref{eqn:pp_eom_les}) is given by $\Sigma$ only, the truncation of the corresponding integrals is possible, and a convergence of the numerical simulations can be obtained for $t_c=50$ inverse hoppings.

\begin{figure}[tbp]
\includegraphics[scale=1]{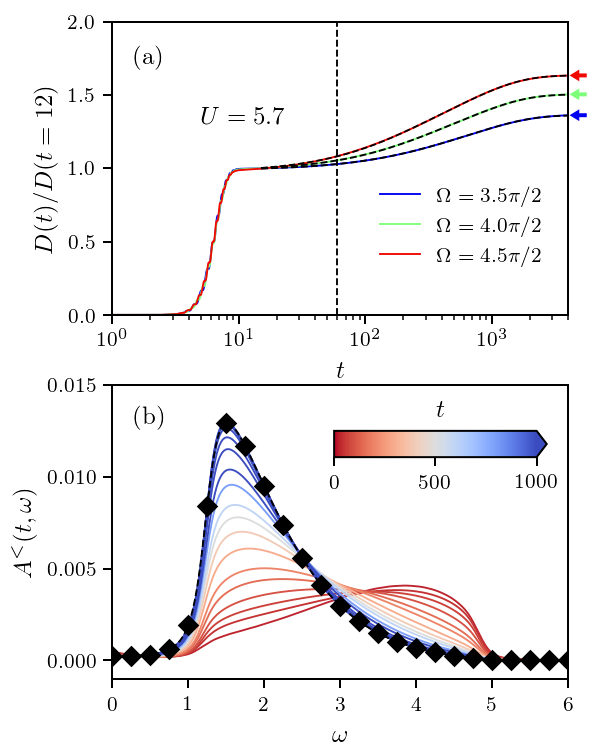}
\caption{\label{fig:NCA_FIG2}(a) The time evolution of the normalized doublon density $D(t)=d(t)-d(0)$ plotted for different frequencies $\Omega$ of the laser pulse. The vertical dashed line indicates the maximum time-scale for untruncated solutions. The horizontal arrows indicate the thermal equilibrium values $D_{th}$ with the same total energy. The amplitude of the pulse in these calculations is adjusted such that immediately after the pulse $D(t=12)$=0.01. The black dashed lines correspond to the fit of a double exponential function $d(t)={D_th}+a\exp{-(t-t_p)/\gamma}+c\exp{-(t-t_p)/\tau}$ with the relaxation times $\tau=671.30,724.94,789.17$ and $\gamma=143.14,164.85,182.92$ (top to bottom). (b) Occupied density of states of the upper Hubbard band for selected times $t$ between 30 and 4000 inverse hoppings (red to blue) and 
the distribution in the final thermalized state  (black diamonds).
}
\end{figure}

To examine the physical properties of the long-time solution we evaluate the time-evolution of the double occupancy $d(t)=\sum_i\expval{n_{i\uparrow}(t) n_{i\downarrow}(t)}$ after excitation of the system with various frequencies $\Omega$ (Fig.~\ref{fig:NCA_FIG2}a). As in Ref.~\onlinecite{Werner2014}, we show the increase of the double occupancy with respect to the initial value, $D(t)=d(t)-d(0)$, normalized to the value after the pulse. We choose an interaction $U=5.7>W$, resulting in a Mott gap which is smaller than the bandwidth, so that impact ionization is possible. The  double occupancy exhibits a sharp increase after the photo-excitation. The initially created high-energy doublons relax via impact ionization, creating additional low-energy doublons. These low-energy doublons then thermalize on even longer timescales. Due to the computational cost for long-time simulations, only the onset of this final thermalization of the state was observed in previous numerical calculations \cite{Werner2011,Werner2014}. Thermalization is confirmed by comparing the final double occupancy to the double occupancy in a thermal state with the same total energy as the photo-excited state, see arrows in Fig.~\ref{fig:NCA_FIG2}(a). The energy of the photo-excited state is obtained from the current $j(t)$ by integrating the electric power $\Delta E_{tot}$=$\int\dd t j(t)E(t)$. Note that the agreement between the thermal equilibrium state and the final state of the propagation within the numerical accuracy proves the conservation of the total energy in the truncated evaluation of the KBE.

The fact that the relaxation is determined by distinct mechanisms on different timescales implies that the time-dependence of the double occupancy does not follow a single exponential function. Instead, one can fit a double exponential function $d(t)=D_{th}+a\exp{-(t-t_p)/\gamma}+c\exp{-(t-t_p)/\tau}$ with separate relaxation times $\gamma$ ($\tau$) for high (low) energy doublons to the evolution of the double occupancy, see Fig.~\ref{fig:NCA_FIG2}(a). The values for $\gamma$ and $\tau$ agree with the previously reported values~\cite{Werner2014}. However, one can now see that this double exponential provides a good description of the relaxation for much longer times: In the present study, the fit is done from $t=15$ to $t=4000$, compared to the maximum time $t=60$ which would be accessible in the untruncated simulation, and which would only include the onset of the thermalization. 

The conversion of high-energy to low energy doublons can also be seen in the occupied density of states in the upper Hubbard band obtained from the partial Fourier transform  
\begin{equation}
A^{<}(t,\omega)=\frac{1}{\pi}{\rm Im}\int\!\! \dd s\, G^{<}(t+s/2,t-s/2)e^{\imag \omega s},
\label{eqn:Wigner_transform}
\end{equation}
plotted in Fig.~\ref{fig:NCA_FIG2}(b). The incident laser pulse initially induces a finite occupation towards the upper edge of the upper Hubbard band. As time progresses, these high-energy doublons decay into low-energy states, transferring spectral weight from the upper band edge to the lower band edge. The role of impact ionization in the initial dynamics can hence be quantified by calculating the ratio of the time-dependent change of the integrated spectral weight at the upper and lower edge. In our calculations, this ratio turns out to be around three, which indicates the decay of a single high-energy doublon (hole) into three low-energy doublons (holes).

\section{\label{sec:FLEX}Dynamics at the superconducting phase transition}

Another situation where one can expect physics to emerge at largely separated timescales is the distinct dynamics of electrons and order parameter fluctuations in the vicinity of a non-equilibrium phase transition. As a paradigmatic example, we  illustrate the use of the truncated KBE for the dynamics of superconducting fluctuations after an interaction quench in the attractive three-dimensional Hubbard model in the vicinity of the superconducting phase transition. While the late-time dynamics of the order parameter should follow a phenemenological description within time-dependent Ginzburg-Landau theory \cite{Lemonik2017}, microscopic simulations of the Hubbard model could demonstrate how early non-thermal electron distributions can leave a signature in the order parameter fluctuations at later times.\cite{Stahl2021}  The consistent description of the intertwined evolution of electrons and order parameter fluctuations  in this case was achieved by a solution of the full KBE  with a momentum dependent self-energy $\Sigma_k$ constructed from the fluctuation exchange approximation (FLEX). This could resolve the initial thermalization of the electronic system and the opening of a pseudo-gap in the one particle spectrum, but could not capture the time propagation to a regime where a time-dependent Ginzburg-Landau Theory would be fully applicable. Motivated by this, we  analyze below to what extent the same setting as in Ref.~\onlinecite{Stahl2021} can be addressed using the truncated KBE. 

The analysis is done for the three-dimensional Hubbard model,
\begin{equation}
H=\sum_{\bm k\sigma}(\epsilon_{\bm k}-\mu) c^\dagger_{\bm k\sigma}c_{\bm k\sigma}+\frac{U}{N}\sum_{\bm q}\Delta_{\bm q}^\dagger\Delta_{\bm  q},
\label{eqn:Hubbard}
\end{equation}
where the interaction term is already written in terms of the superconducting order parameter  $\Delta_{\bm q}=\sum_{\bm k}c_{\bm k\uparrow}c_{-\bm k+\bm q\downarrow}$ and $\Delta_{\bm q}^\dagger=\sum_{\bm k} c^\dagger_{-\bm k+\bm q\downarrow}c^\dagger_{\bm k\uparrow}$ for an attractive interaction ($U<0$). For the numerical simulation we assume a continuum limit (electrons in the vicinity of a band minimum), so that the dispersion is $\epsilon_{\bm k}=k^2$, and momentum sums become $(1/N)\sum_{\bm k}=\int d^3k/(2\pi)^3$, with a large momentum cutoff $|\bm  k|<k_c$.  We choose the cutoff $k_c=\pi$ and $\mu=2.59$, so that $k_F=0.57k_c$, and  approximately $18\%$ of the states within the cutoff are filled. We have confirmed that the cutoff is large enough so that resulting errors, such as a violation of the density conservation, are negligible.
The solution in terms of FLEX~\cite{Bickers1989} combines 
 the random phase approximation (RPA) for the superconducting fluctuations $\chi_q(t,t')=-\imag/N\expval{T_{\mathcal{C}}[\Delta_q(t)\Delta^\dagger_q(t')]}$, given by 
\begin{equation}
\chi_q(t,t')=\chi^0_q(t,t')+\int_{\mathcal{C}}\!\!\! \dd \bar{t}\,\,\chi^0_q(t,\bar{t})U(\bar{t})\chi_q(\bar{t},t')
\label{eqn:RPA}
\end{equation}
in terms of the bare propagator $\chi_q^0(t,t')=\imag/N\sum_kG_k(t,t')G_{q-k}(t,t')$, and  the KBE \eqref{eqn:KBEretarded}$ - $\eqref{eqn:KBElesser}
for the Green's function $G_k(t,t')$, with a fully self-consistent FLEX self-energy 
\begin{equation}
\Sigma_k(t,t')=-\frac{\imag}{N}\sum_q U(t)\chi_q(t,t')U(t')G_{q-k}(t',t).
\end{equation}
This set of equations can be efficiently implemented for a spherical symmetric system, because $G_k$, $\Sigma_k$ and $\chi_q$ all depend only on the absolute value of $k$, which makes the simulation of a three-dimensional system feasible on 400 $k$-points [see  Ref.~\onlinecite{Stahl2021} for details]. 
The dynamic is initiated by a sudden quench of the interaction $U$ from the initial paramagnetic equilibrium state ($U=-3$, $T=0.11$) to $U=-3.5$, which is associated with the superconducting phase of the system in equilibrium. As discussed in Ref.~\onlinecite{Stahl2021}, a gap opens after the quench due to the enhanced fluctuations as the precursor of a phase transition.

\begin{figure}
\includegraphics[scale=1]{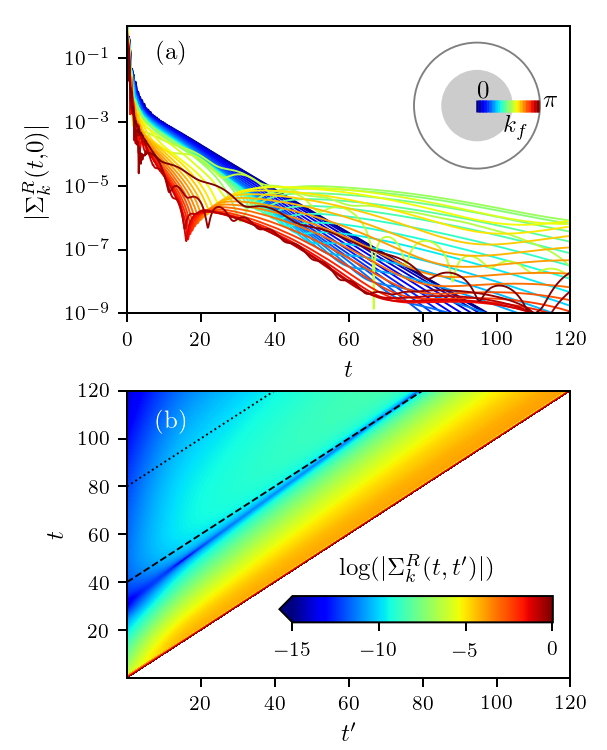}
\caption{(a) Absolute value of the retarded component of the self energy $\Sigma_k$ for different $k$. The color code of the values of $k$ is indicated by the circle on the right; the inner circle represents occupied states below the Fermi-momentum $k_f$. 
(b) Absolute value of $\Sigma_k$ for $k=1.65$ (least decaying value in (a)), slightly below the Fermi-edge as function of $t$ and $t'$.
The dashed (dotted) line indicates the cutoff $t_c=40$ ($t_c=80$) used in the simulations below.
}
\label{fig:FLEX_SIGMA}
\end{figure}
\begin{figure}
\includegraphics[scale=1]{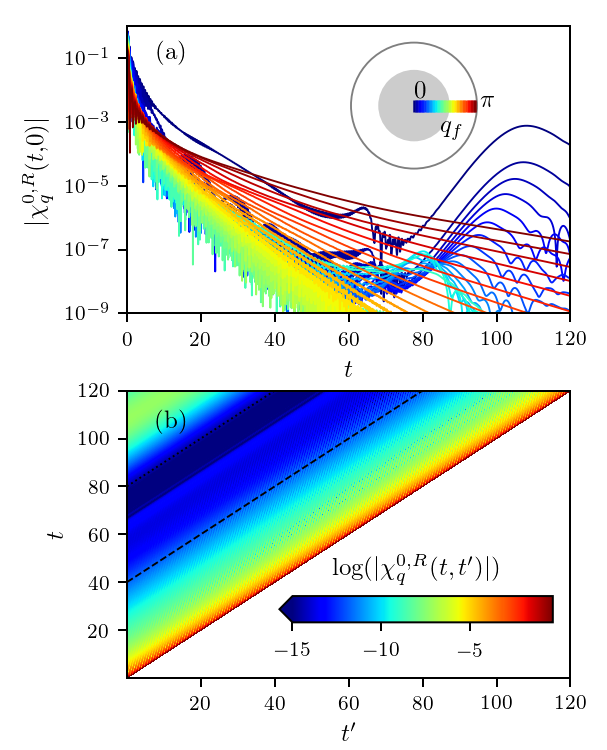}
\caption{(a) Absolute value of the retarded component of the bare propagator $\chi^0_q$ for different $q$. The color key for the momenta is shown by the circle on the right.
(b) Absolute value of $\chi^0_q$ for $q=0$ as a function of $t$ and $t'$. The dashed (dotted) line indicates the cutoff $t_c=40$ ($t_c=80$).
}
\label{fig:FLEX_Chi}
\end{figure}
\begin{figure}
\includegraphics[scale=1]{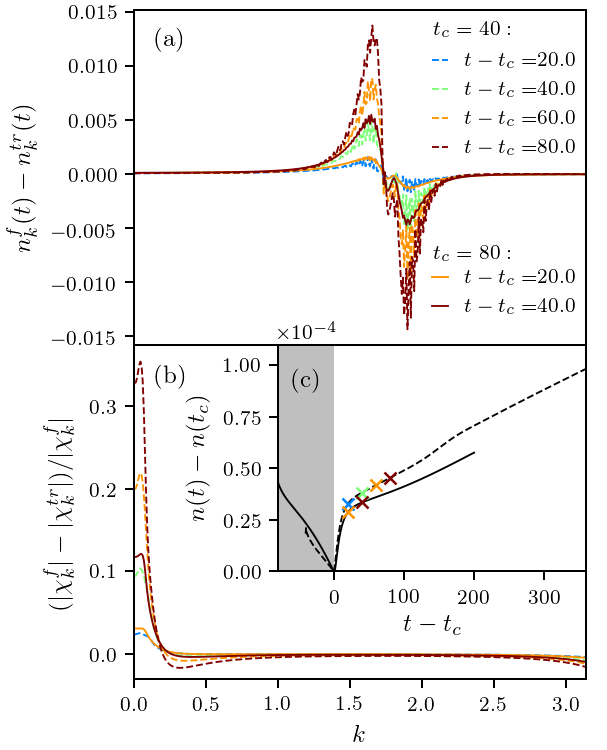}
\caption{(a) Deviation of the distribution $n_k^{tr}(t)=\imag G_k^<(t,t)$, as obtained in the truncated KBE, from the fully propagated distribution $n_k^{f}$ for different truncation times $t_c$ (solid, dashed) and times (color). 
(b) Relative deviation of the fluctuations $\chi^{tr}_k$, obtained from the the truncated KBE, from the fully propagated $\chi^{f}_k$ for the same $t_c$ and $t$ as in (a).
(c) Particle density $n(t)$ normalized for the density $n(t_c)$ against the time relative to the truncation for different $t_c$ (solid,dashed). Colored marks indicate the time steps shown in (a) and (b).}
\label{fig:FLEX_ERROR}
\end{figure}

Using the untruncated KBE we could simulate the FLEX equations up to $t=120$  hopping times.
A closer look at the $\Sigma_k(t,t')$ in Fig. \ref{fig:FLEX_SIGMA} shows a general decay of the retarded component for the individual  momenta in the full propagation. The slowest decaying mode of Fig.~\ref{fig:FLEX_SIGMA}(a), which corresponds to the momentum  closest to the Fermi-surface, is plotted as a function of $t$ and $t'$ in Fig.~\ref{fig:FLEX_SIGMA}(b), indicating that a truncation with $t_c=40$ ($t_c=80$) neglects values below a threshold of $\mathcal{O}(10^{-5})$.

As the FLEX approximation is summing over diagrams up to infinite order in $U$ we need to consider the internal time integrals in the  self-energy, which are all embedded in the RPA equation~\eqref{eqn:RPA}. The RPA equation for the susceptibility is equivalent to the integral equation \eqref{vie2} introduced in Sec.~\ref{sec:TruncKBE}, where $\chi$ corresponds to the function $G$, and $\chi^0$ represents the integral kernel $K$ which should decay as a function of relative time for a truncation scheme to be applicable.
The superconducting  susceptibility, which is related to the retarded component of $\chi$, will develop a singularity at $\omega=0$ at the superconducting phase transition, which translates to a constant value in the time domain. The function $\chi$ can therefore be expected to decay slowly as a function of relative time.
The bare susceptibility $\chi^0$ on the other hand is moderately decaying, apart from the $q=0$ component which shows a revival dynamics after an initial decay in Fig.~\ref{fig:FLEX_Chi}. Fortunately, the $q=0$ component does not enter the self-energy $\Sigma$ as the corresponding integral weight is proportional to $q$~\cite{Stahl2021} and higher momenta show a significantly less pronounced revival, yielding a threshold of $\mathcal{O}(10^{-4})$ for the chosen cutoff times.    

\begin{figure}
\includegraphics[scale=1]{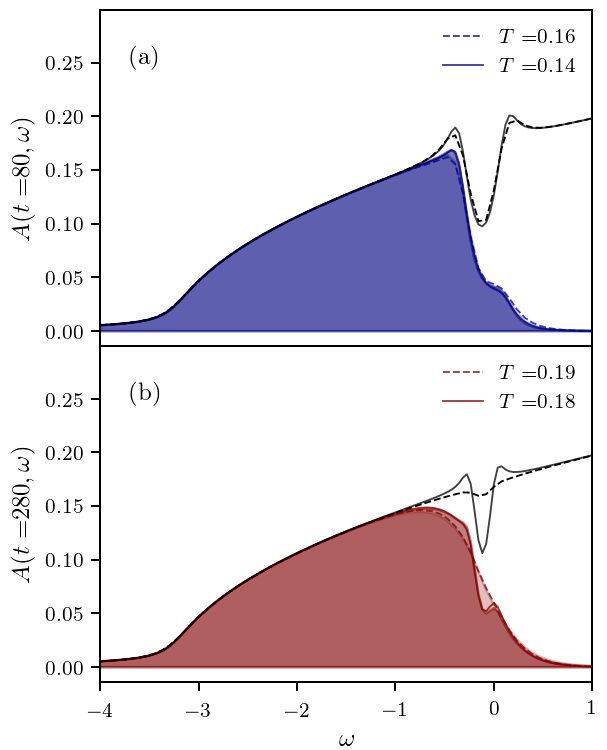}
\caption{(a) Spectral function (black) obtained from the local retarded Green's function $(-G^R)$ by the Wigner transformation defined in Eq.~\eqref{eqn:Wigner_transform} and the occupied spectrum obtained from Eq.~\eqref{eqn:Wigner_transform} (blue filling) for $t_c=40$ (dashed) and $t_c=80$ (solid) at $t=80$. Blue lines correspond to the best approximation of the occupied spectrum by a thermal state according to Eq.~\eqref{eqn:FDT} with temperature $T$. (b) Same quantities for $t=280$.}
\label{fig:FLEX_TEMP}
\end{figure}

Despite the apparently small value of the integral kernels $\Sigma_k$ and $\chi^0_q$ at relative times beyond a cutoff time $t_c=40$ or even $t_c=80$, one finds that the solution of the full coupled equations is still sensitive to the truncation error.
First of all, one finds a small  ($10^{-4}$) increase of the particle number after $t=t_c$, when going from the full to the truncated KBE, see Fig.~\ref{fig:FLEX_ERROR}(c). This increase after $t_c$ is comparable to the error on the particle number during the full propagation scheme on 80 inverse hoppings, which was identified as an artifact of the momentum cutoff $k_c$ in Ref.~\onlinecite{Stahl2021}. 
After the sudden jump in the particle density, the particle number continues to grow in the truncated scheme with a slope similar to the initial decrease observed in the untruncated propagation. 
A comparison of the electronic distribution in the truncated and untruncated propagation scheme reveals that the increase of particles happens dominantly for states above the Fermi-edge, and therefore increases the energy of the system, see Fig.~\ref{fig:FLEX_ERROR}(a). The modification of the particle density is further accompanied by a significant decrease of the superconducting fluctuations, see Fig.~\ref{fig:FLEX_ERROR}(b). 

The lower value of the superconducting fluctuations in the truncated evolution can be understood by evaluating the 
electronic spectral function and distribution function
in terms of the  Wigner transformation~\eqref{eqn:Wigner_transform}, see Fig.~\ref{fig:FLEX_TEMP}. 
Since the spectrum yields a slow redistribution of the particles at the Fermi-edge for $t\geq80$, the electronic distribution can be approximated by thermal states, which fulfill the fluctuation dissipation theorem:
\begin{equation}
A^<(t,\omega)=-\frac{2}{\pi}\,\text{Im}\,G^{R}(t,\omega)f(T,\omega),
\label{eqn:FDT}
\end{equation}
where $f(T,\omega)$ is the Fermi function.
The obtained electronic temperature $T$ increases with the propagation time, causing the decrease of the fluctuations, which leads ultimately to the closure of the pseudo-gap. As the increase of the temperature depends on $t_c$, the heating can be identified as an artifact of the truncation, instead of a feature of the non-equilibrium physics.

This analysis indicates that the truncation error behaves empirically like an additional thermal bath attached to the system, which erases memory longer than $t_c$ due to thermal fluctuations and leads to artificial heating. A direct mapping of the effect onto a specific physical bath is, however, not possible.

\section{\label{sec:Conclusion}Conclusion}
We studied  the effect of memory truncation in the KBE 
within different diagrammatic approximations for the Hubbard model, including both DMFT in the weak and strong coupling limit, and the non-local FLEX approximation. The approach exploits the decay of the self-energy with relative time $|t-t'|$, and truncates the memory integrals in the KBE beyond a cutoff time $t_c$. With this, the numerical effort to compute the Green's function on the domain $|t-t'|\leq t_c$ on an equidistant time grid $t=n\dt$ reduces from $\mathcal{O}(n^3)$ to $\mathcal{O}(nn_c^2)$, and the required memory reduced from $\mathcal{O}(n^2)$ to $\mathcal{O}(n_c^2)$, eliminating the previously limiting memory bottleneck in long time propagations ($n_c=t_c/\dt$).

The minimal cutoff depends strongly on the physical problem. We find that within the benchmark applications of non-equilibrium DMFT to the single band Hubbard model the results can be converged by increasing the cutoff $t_c$, and simulations are possible to times $t$ which are orders of magnitude longer than $t_c$ itself, and thus orders of magnitude longer than what would be possible with the full KBE. We have demonstrated this with paradigmatic problems which show largely separated timescales, i.e, (i), thermalization and prethermalization after an interaction quench in the Hubbard model at weak coupling, and (ii), impact ionization and thermalization in a photo-excited Mott insulator.

In the case of DMFT, the integral kernels which control the cutoff are all spatially local quantities, i.e., the local self-energy, and the impurity hybridization function in the case of the strong-coupling impurity solver. The truncation of the memory integrals turned out to be more subtle in the case of the FLEX simulation, which requires the joint solution of RPA equations for the momentum dependent  pairing fluctuations $\chi_q$  and the Dyson equation with a momentum dependent self-energy $\Sigma_k$. In this case, the decay of the relevant integral kernels depends strongly on momentum. While this prevented to simulate the evolution to times much longer than with the full KBE in the present work, the analysis shows a possible way forward: First of all, one finds that by implementing a suitable momentum dependent cutoff $t_c(k)$, Dyson and RPA equations for most momenta could be solved efficiently (with a shorter cutoff), while long cutoff times are only needed for the determination of the Green's function with momenta close to the Fermi surface, and for $\chi_q$ with momenta $q$ close to the pairing instability $q=0$. Moreover, empirically we note that the truncation error manifests itself in a similar way as an additional heat bath. Conversely, one can expect that including real physical thermal reservoirs to the model may allow for shorter truncation times.  For the long-time simulation of the dynamics in condensed matter, the electronic subsystem anyway cannot be considered as isolated, and such thermal reservoirs should be incorporated, e.g., to represent the coupling to phonons. 
 
The momentum-dependent cutoff could eventually allow for true multi-scale simulations of the condensed matter dynamics, with a consistent treatment of the non-thermal electron dynamics on femtosecond timescales and the order parameter dynamics on picosecond timescales. Further possible future developments include similar memory truncation to higher-order self-energy approximations within the strong coupling solution of the DMFT impurity model \cite{eckstein2010} (such as the one-crossing approximation), and the combination of the memory truncation scheme with compact basis representations of the non-equilibrium Green's functions \cite{Kaye2021}.

\begin{acknowledgements}

We thank Francesco Grandi and Denis Gole\v z for useful discussions. 
We acknowledge financial support from the DFG Project 310335100 and the ERC starting grant No.~716648. JL acknowledges the Marie Skłodowska Curie grant agreement No 884104 (PSI-FELLOW-III-3i). P. W. acknowledges support from ERC Consolidator Grant No. 724103. The numerical calculations have been performed at the RRZE of the University Erlangen-Nuremberg. 
\end{acknowledgements}


\end{document}